\documentclass[]{aastex62}
\usepackage{physics}

\def\itemrange#1{%
\addtocounter{enumi}{1}%
\edef\labelenumi{\theenumi--\noexpand\theenumi.}%
\addtocounter{enumi}{-1}%
\addtocounter{enumi}{#1}%
\item
\def\labelenumi{\theenumi.}}

\begin{document}

\title{How Much SETI Has Been Done? Finding Needles in the $n$-Dimensional Cosmic Haystack}
\author[0000-0001-6160-5888]{Jason T.\ Wright}
\affiliation{Center for Exoplanets and Habitable Worlds, Department of Astronomy \& Astrophysics, The Pennsylvania State University, 525 Davey Laboratory, University Park, PA 16802, USA}

\author[0000-0001-8401-4300]{Shubham Kanodia}
\affiliation{Center for Exoplanets and Habitable Worlds, Department of Astronomy \& Astrophysics, The Pennsylvania State University, 525 Davey Laboratory, University Park, PA 16802, USA}

\author[0000-0003-0790-7492]{Emily Lubar}
\affiliation{Center for Exoplanets and Habitable Worlds, Department of Astronomy \& Astrophysics, The Pennsylvania State University, 525 Davey Laboratory, University Park, PA 16802, USA}

\correspondingauthor{Jason T.~Wright}
\email{astrowright@gmail.com}

\begin{abstract}
Many articulations of the Fermi Paradox have as a premise, implicitly or explicitly, that humanity has searched for signs of extraterrestrial radio transmissions and concluded that there are few or no obvious ones to be found. \citet{Tarter10} and others have argued strongly to the contrary: bright and obvious radio beacons might be quite common in the sky, but we would not know it yet because our search completeness to date is so low, akin to having searched a drinking glass's worth of seawater for evidence of fish in all of Earth's oceans. Here, we develop the metaphor of the multidimensional ``Cosmic Haystack'' through which SETI hunts for alien ``needles'' into a quantitative, eight-dimensional model, and perform an analytic integral to compute the fraction of this haystack that several large radio SETI programs have collectively examined. Although this model haystack has many qualitative differences from the \citet{Tarter10} haystack, we conclude that the fraction of it searched to date is also very small: similar to the ratio of the volume of a large hot tub or small swimming pool to that of the Earth's oceans. With this article, we provide a Python script to calculate haystack volumes for future searches and for similar haystacks with different boundaries. We hope this formalism will aid in the development of a common parameter space for the computation of upper limits and completeness fractions of search programs for radio and other technosignatures.
\end{abstract}

\keywords{extraterrestrial intelligence}

\section{How Much Searching Have We Done?} \label{sec:intro}

\subsection{The Fermi Paradox and the supposed ``silence'' of the Galaxy}
The so-called\footnote{It is ``so-called'' because, as \citet{FermiParadox} has documented, it ``is neither Fermi's nor a paradox.''  See, however, \citet{CirkovicFermi}.} ``Fermi Paradox'' is usually used as shorthand for the supposed inconsistency between optimistic estimates of the number of intelligent species in the Galaxy according to the Drake Equation \citep{DrakeEquation} and the fact that we have not yet discovered any of them. 

A particularly trenchant formulation of this inconsistency is that of \citet{hart75}, who dubs the fact that there are not aliens on Earth now ``Fact A.''  This fact is supposedly surprising because the time for an exponentially growing population to travel to every star in the Galaxy under even rather conservative assumptions of the feasibility of interstellar travel is short compared to the age of the Galaxy \citep[e.g.][and references therein]{GHAT1}. Many have used this and similar articulations of the Fermi Paradox to argue that the entire endeavour of SETI is a ``waste of time'' \citep[e.g.][]{Tipler93}.

One of the central assumptions behind the significance of ``Fact A'' to SETI is that any spacecraft arriving in the Solar System would result in the permanent settlement of the Earth.  \citet{GHAT1} noted that his assumption fails for the only technological species we have an example of: humanity. The ratio of the timescales for humans to cross the Earth to the age of the technology that enables such travel is small: today for aircraft, it is roughly $10^{-4}$ (days over decades), but even a thousand years ago it was under $10^{-3}$ (years over millennia). Despite this, most spots on Earth show no obvious signs of humans or our technology. If we analogously relax Fact A and assert merely that there are no alien machines in the Solar System today, then its truth becomes quite uncertain because we have not done much searching for such machines \citep[e.g.][]{Haqq12,Davies12}.

Other variants on ``Fact A'' are the ``Great Silence'' of \citet{Brin83} and the ``Eerie Silence'' of \citet{Davies_Eerie}, both of which refer to the more general fact that we have not yet found any evidence of extraterrestrial intelligent life in the universe. The term ``silence'' in the these expressions is sometimes interpreted to mean that we have established that there are no obvious radio or other communicative signals to be found in space---i.e. that we have conducted a thorough search for signals, and concluded that there are very few or none of them. 

However, this is not the case. Despite the passage of nearly 60 years since the first radio SETI searches, very little actual searching has been done compared to the amount needed to rule out the presence of a even a large number of ``loud'' beacons \citep[e.g.][]{Shostak14}. Perceptions to the contrary abound, as many authors use the ``failure'' of SETI to date as evidence that we should stop looking \citep[e.g.][]{Tipler93}, that we need to move to new search strategies \citep[e.g.][]{Davies_Eerie}, or that we should re-evaluate the fundamental assumptions of the field \citep[e.g.][]{Brin14}.

Quantifying the amount of searching that has been done should help to dispel this misconception. \citet{Tarter10}, for instance, have estimated that ruling out the existence of radio beacons based on the amount of searching to date is akin to basing one's opinion on the existence of marine life on dipping a single drinking glass into the ocean and seeing if it contains a fish.

\subsection{Quantifying Upper Limits in SETI}

Searches for first examples of hypothesized physical phenomena  proceed through a series of progressively stringent null results.  The thoroughness and speed of a search are typically quantified by parameterizing the search space and calculating the volume explored. Thus, when a well-designed and -executed program ``fails'' to find the sought-after phenomenon, that null result can be expressed as the lower surface of the explored volume, which represents a new upper limit that future experiments can improve upon.  Once the entire parameter space a phenomenon could occupy has been excluded by experiment, we can speak of the null result as being dispositive: the sought-after phenomenon does not exist \citep{Wang12}.

However, at the beginning of a search, typical upper limits are far from dispositive and can even seem absurdly high. For instance, the first upper limits on the flux of solar neutrinos were much higher than the expected flux, and indeed a reviewer of the paper describing the experiment amusingly commented:
\begin{quote}
Any experiment such as this, which does not have the requisite sensitivity, really has no bearing on the question of the existence of neutrinos. To illustrate my point, one would not write a scientific paper describing an experiment in which an experimenter stood on a mountain and reached for the moon, and concluded that the moon was more than eight feet from the top of the mountain. \citep[][p. 245]{neutrinos}
\end{quote}
But the reviewer is incorrect. It is not at all hard to imagine that prehistoric humans around the world performed just such an experiment, told others about it, and in so doing significantly advanced their understanding of the cosmos. 

Similarly, in SETI the first upper limits will necessarily rule out only extremely obvious manifestations of alien technology, and may seem at first to ``have no bearing on the question of the existence of'' alien life. For instance, \citet{GHAT3} put weak upper limits on the total waste heat across entire galaxies, concluding that out of 100,000 galaxies, none had a galaxy-spanning species harnessing more than 85\% of the galaxy's starlight. This result might hardly be surprising given that this upper limit is near the extreme limit of what galaxy-spanning technology could accomplish; indeed, contra \citet[p.~58]{cirkovic18} it certainly does not ``tremendously limit the parameter space describing very advanced civilizations.'' Nonetheless, it represents a necessary first step toward more stringent and meaningful upper limits on Type {\sc iii} Kardashev species \citep{kardashev64}.

Unlike solar neutrinos, however, the potential number of technosignatures\footnote{By analogy with ``biosignatures,'' ``technosignatures'' are the signatures of alien technology. The term apparently originates with \citet{Tarter06}.} is quite large and difficult to parameterize.  In one early visualization of the search space for radio communication SETI, \citet{Wolfe81} identified several relevant parameters: polarization, repetition rate of transmission, signal modulation, three spatial dimensions, central transmission frequency, and receiver sensitivity. They then ``compressed'' the three dimensions of polarization, frequency of transmission, and signal modulation by considering only obvious, continuous signals searched for with receivers sensitive to orthogonal polarizations. Next, they compressed the three spatial dimensions into a single axis in terms of the number of targets or beams surveyed. They then graphically illustrated a three-dimensional ``Cosmic Haystack''\footnote{The earliest attestation of the term ``Cosmic Haystack'' in print we are aware of is \citet{CosmicHaystack}. In the mid-1990s the term began to appear in non-SETI astronomical contexts \citep[e.g.][]{Werner95,Grenier00}.} through which various SETI programs had hunted, looking for ``needles.''\footnote{``To look for a needle in a haystack,'' meaning to search for something very difficult to find, is often misattributed to Cervantes because it appears in a 1703 English translation of {\it Don Quixote} by Peter Anthony Motteux, but the Spanish original actually reads ``buscar\ldots a Marica por R\'avena,'' i.e.\ to look for a particular ``Maria'' in a place where that is a very common name \citep{Lathrop06}.  The earliest attestations to the phrase and its variants in the \citet{needle} are from Saint Thomas More in 1557 (``To seke out one lyne in all hys bookes wer to go looke a nedle in a medow''), Robert Greene in 1592 (``He\ldots gropeth in the dark to find a needle in a bottle [bundle] of hay''), and W.~Rogers in 1779 (``But agreeably to the old adage it was similar to looking for needles in a hay stack.'') Being a common English idiom, the metaphor naturally appears frequently in the astronomical literature \citep[e.g.][]{Raphael45}.} The term and visualization also appear in \citet{Tarter85}.

Similarly, \citet{papa85} described the ``SETI search space'' of two NASA SETI search programs for narrowband transmission in terms of the frequencies searched, fraction of sky surveyed, and instrument sensitivity. He distinguished between the {\it Targeted} search that parametrized the haystack in terms of the number of targets surveyed (discrete sources at good sensitivity) and a {\it Sky Survey} program parameterized in terms of sky coverage (i.e.\ a shallower search for any transmitters in the beam). 
\citet{Tarter06} and \citet{Tarter10} described the search space as a ``nine-dimensional haystack'' in terms of three spatial dimensions, one temporal dimension (when the signal is ``on''), two polarization dimensions, central frequency, sensitivity, and modulation.

\subsection{Extending the Haystack Metaphor}

This ``needle in a haystack'' metaphor is especially appropriate in a SETI context because it emphasizes the vastness of the space to be searched, and it nicely captures how we seek an obvious product of intelligence and technology amidst a much larger set of purely natural products. SETI optimists hope that there are many alien needles to be found, presumably reducing the time to find the first one.  Note that in this metaphor the needles are the {\it detectable signatures} of alien technology, meaning that a single alien species might be represented by many needles.

Much of the SETI literature is dedicated to guessing or reasoning which haystacks might contain many needles. For instance, if we posit that extraterrestrial species are trying to get our attention and so  have created obvious ``beacons'' \citep[e.g.][]{Dixon73}, then we can focus our efforts on those haystacks or parts of haystacks where needles would most obviously be left for us to find \citep[i.e.~we can identify which haystacks represent Schelling points in a game of mutual search, for instance by guessing at ``magic frequencies;''][]{WrightExoplanetsSETI}.

There is also the issue of recognizing a needle when one finds it, and not mistaking it for a particularly odd piece of straw. Communication SETI typically seeks signals so obviously artificial that there will be no doubt that the search has succeeded. Artifact SETI seeks astrophysical anomalies and exotica that are consistent with predictions for alien technology, but may not be dispositive of such technology \citep{GHAT1}. Here the metaphor breaks down somewhat, but we might interpret the needles and straw in these haystacks to be hard to distinguish. Finding such maybe-needles nonetheless helps communication SETI efforts to narrow down which parts of their haystack to focus on to find more unambiguous needles. 

\subsection{Survey  Figures of Merit}
\label{SSFM}
A given telescope and instrumentation will take some amount of time to complete a survey for a given sort of technosignature down to a given sensitivity. Intuitively, one would like to define a {\it survey speed}---i.e.\ haystack volume surveyed per unit of time---that described the haystack volume searched as a linearly increasing function of integration time or telescope time.\footnote{In practice, this will not be strictly possible, because different parts of a haystack will be searched at different powers of time; also, some survey speeds are calculated per pointing, ignoring the integration time required.}

\citet{Drake84} and \citet{Enriquez17} provide good discussions of some proposed figures of merit for ``survey speed'' in radio SETI work, and describe the difficulties in constructing one that allows fair comparison among surveys with different strategies and assumptions. Such figures of merit are usually of smaller dimensionality than haystacks because many dimensions are common across surveys, and so add no new information when comparing speeds of surveys. For instance: one rarely considers the sensitivity to polarization of the instrument because modern radio instrumentation is equally sensitive to orthogonal polarizations; one would not consider the repetition rate of the transmitter because searching for lower repetition rates simply requires more telescope time regardless of the instrumentation used.

As a more specific example, \citet[][p.~4]{Drake84} described a commonly used figure of merit for narrowband radio searches:
\begin{equation}
\label{DrakeFOM}
{\rm BW}_i \Omega \phi_{\rm min}^{-\frac{3}{2}}
\end{equation}
\noindent where ${\rm BW}_i$ is the instrument bandwidth; $\Omega$ is the fraction (solid angle) of the sky surveyed, and $\phi_{\rm min}$ is the minimum detectable flux (W/m$^2$) in a single observation. The $-3/2$ dependence arises because the distance $d$ out to which one can detect a signal of a given intrinsic strength scales as $d \propto \phi_{\rm min}^{-\frac{1}{2}}$, and the volume one probes in a single beam scales as $\Omega d^3$. The ``haystack'' implied by the Drake figure of merit is thus a literal 3D volume times a frequency dimension, hence 4D. \citet{Lesh15} extend this haystack in two more dimensions, incorporating the polarization and number of stars surveyed as additional parameters. 

As pointed out by \citet{Enriquez17}, such haystacks have some curious properties, as a search speed, that seem counter-intuitive (but which we will see later can be virtues.)  For instance, it treats all points in space equally and does not give a higher weight for regions (directions) more likely to host signals, such as (for instance) nearby stars. It also considers signal strength in a rough way: within a distance $d$ it gives equal weight to all space within the beam (even though one's sensitivity to very nearby sources is much better, so one's survey speed there should be in some sense higher), and zero weight to the space beyond $d$ (even though one's sensitivity to signals beyond this distance is nonzero). The primary value of the Drake figure of merit, then, is that it offers a scaling relation useful for comparing the speeds of radio surveys hunting for narrowband emission anywhere in the sky down to a certain {\it transmission} strength.

\citet{Tarter01} describes the figure of merit proposed by \citet{Dreher97} which is a function of the Equivalent Isotropically Radiated Power (EIRP)\footnote{EIRP is a standard radio measure of artificial transmitter luminosity, calculated from the flux $\phi$ and distance $d$ of a source as ${\rm EIRP} = 4\pi d^2\phi$.  It is equal to the total power of a transmission multiplied by the gain of the transmission antenna (which generally will not be known for alien transmitters, even in the event of a detection, because it is a property of the transmitter itself, not necessarily its signal).} of a hypothetical transmitter, given by 
\begin{equation}
\ln\left(\frac{\nu_{\rm high}}{\nu_{\rm low}}\right)N_{\rm stars}(\rm EIRP)  \eta_{\rm pol} N_{\rm looks}
\end{equation}
\noindent where $N_{\rm stars}$ is the number of stars in a given pointing to which one can detect a signal of strength EIRP, $N_{\rm looks}$ accounts for the number of visits to the same sky position, $\eta_{\rm pol}$ is a number between 0 and 1 that describes the survey sensitivity to polarization, and the $\nu$ terms describe the lowest and highest frequencies surveyed. We note that, in our terms and notation, two surveys surveying the same parts of the sky at relatively narrow bandwidths will compare by this metric roughly as
\begin{equation}
\frac{{\rm BW}_i}{\nu}\Omega\phi_{\rm min}^{-\frac{3}{2}}N_{\rm looks}
\end{equation}
where we have approximated the number of stars surveyed by the volume of space explored out to the limiting distance given by the survey sensitivity. Note that this recasting breaks down for surveys so sensitive or complete that they ``run out'' of stars in the Galaxy \citep[see Figure 4 of][]{Tarter01}.

\citet[][eq.~9]{Enriquez17} prefer a metric proportional to the inverse square of the minimum detectable flux of an unresolved narrowband signal, which they call the {\it Survey Speed Figure of Merit}:
\begin{equation}
{\rm BW}_i \phi_{\rm min}^{-2}
\end{equation}
\noindent appropriate for the Breakthrough Listen survey \citep{BL}. The motivation here is that, for observations of fixed channel bandwidth (i.e.~the bandwidth of the widest resolvable signal), sensitivity to signals of a given strength scales as the usual $\sqrt{\tau}$ where $\tau$ is the observation or integration time, so this metric is linear in time. That is, a project with a figure of merit twice as large will complete the same survey in half the time. Solid angle does not appear in this metric because Breakthrough Listen is a targeted survey performing observations down to a given {\it instrumental} sensitivity, and so the survey is defined by the number of stars it surveys, independent of their distance, not by a physical volume. To that end, \citet{Enriquez17} also define the {\it Continuous Wave Transmitter Figure of Merit} of an entire survey, inversely proportional to  
\begin{equation}
\frac{{\rm BW}_i}{\nu} N_{\rm stars} {\rm EIRP_{min}}^{-1}  
\end{equation}

\noindent where $\nu$ is the central frequency of the instrument bandpass and EIRP$_{\rm min}$ is the EIRP of the weakest transmitter that could be detected. Because it is based on the minimum detectable EIRP and not detectable flux, this figure considers surveys of nearby stars more meritorious than surveys of an equal number of more distant stars (because the former can detect weaker signals).

The figure of merit one chooses thus depends on the haystack one is searching and how one is searching it; because different searches historically have had very different search methodologies---even among narrowband radio searches at the same wavelengths---it can be difficult to compare their relative search speeds.  This will also be true with haystacks in general: the size, shape, and dimensionality of the haystack---and one's completeness in searching it---will necessarily depend on the sorts of needles one is searching for.

\subsection{Uniform Distribution of Bright Beacons}

One way to compress the search space that also serves to compare searches is to choose a fiducial transmitter and  follow the Drake figure of merit to determine the upper limit on the uniform space density of such transmitters. That is, we express a search in terms of the number density $n$ of transmitters uniformly or randomly distributed in space and continuously emitting a narrowband signal at fiducial frequency and power. This is not a well-motivated haystack, in the sense that we have no reason to think such a network of uniformly distributed interstellar transmitters would exist,\footnote{But see \citet{Gertz18}.} but it provides an easy-to-understand and one-dimensional metric of search completeness. The number of such transmitters that a search will detect after surveying a solid angle $\Omega$ at sensitivity $\phi_{\rm min}^{-1}$ is given by
\begin{equation}
\frac{n\Omega}{3} d_{\rm max}^3 
\end{equation}
\noindent where $d_{\rm max}$ is the distance of the most distant transmitter that can be detected,
\begin{equation}
d_{\rm max} = \left( \frac{\rm EIRP}{4\pi \phi_{\rm min}}\right)^\frac{1}{2}.
\end{equation}

For instance, the \citet{Enriquez17} L-band search had a sensitivity of 17 Jy in 2.7 Hz channels, and a signal-to-noise ratio detection threshold of 25. For a transmitter with power 1 MW (because our fiducial transmitters are isotropic, this is also the EIRP), this corresponds to a distance limit of $d_{\rm max} = 2.7$ pc. The search covered 10.6 square degrees, or $\Omega = 3.23\times 10^{-3}$ sr.  Since it found no such transmitter, we can put a rough lower limit on the mean separation between any transmitters in L band as
\begin{equation}
l = n^{-\frac{1}{3}} \gtrapprox \left(\frac{\Omega}{3}\right)^\frac{1}{3} d_{\rm max} = 0.27 {\rm pc}
\end{equation}

This is not a very tight limit, but that is not surprising because Breakthrough Listen is a targeted survey performed with a high-gain antenna that measures completeness by the number of nearby stars it observes, not the volume of space it explores.  It does underscore a point about the amount of searching done to date, however, in that a null result such as that of \citet{Enriquez17} can not rule out even very dense networks of continuous megawatt omnidirectional transmitters.\footnote{The haystack that was completely searched by \citet{Enriquez17} is continuous transmitters near their 692 target stars emitting narrowband signals arriving at Earth at the time of their observations in the 1.1--1.9 GHz region with EIRP$\gtrapprox 5.2\times10^{12}$W.}

\section{General Radio SETI Haystacks}
\subsection{The Nine Dimensions of Radio Communication SETI}

\label{9D}
As we have seen, the choice of the number of dimensions for a haystack depends on the method used, and for many surveys, multiple dimensions can be ``collapsed'' into a single dimension, simplifying the problem. For our present purposes, we choose the following (not necessarily orthogonal) dimensions of a radio communication SETI haystack:

\begin{enumerate}
    \item Sensitivity to transmitted or received power
    \item Transmission central frequency
    \itemrange{2} Distance and position (three spatial dimensions)
	\item Transmission bandwidth
    \item Time / repetition rate (i.e.~when the signal arrives)
	\item Polarization
	\item Modulation
\end{enumerate}

As we have seen, the first dimension differs among surveys that seek completeness down to a specific transmission power or a specific received flux; if the former, then there is some non-orthogonality with the distance dimension. 

For targeted surveys, dimensions 3--5 can be collapsed into a single dimension (number of targets surveyed, with distance being folded into the first dimension, if necessary).

Transmission bandwidth has not often been considered in the haystack context, except in the extreme cases of the physical limitations of narrowband emission \citep{Cordes1997} or very broadband pulses \citep{Gindilis73}. We introduce this dimension here to put both modulations onto a single axis, and to account for all of the intermediate cases. We note that, because very short pulses are inherently broadband, there is non-orthogonality here with the time and frequency axes. 

The time-of-observation dimension is complex.  For continuous signals (or ones with such high repetition rates that they are effectively continuous), one does not worry about the time of observation. However if one wishes to detect or rule out not just continuous but also sporadic signals, then the time of observation can matter a great deal. The Wow! signal \citep{Wow} captures this consideration well: failure to confirm the signal could be because it was spurious, or because the signal is intermittent and we have not spent enough time observing it \citep{Gray02}.\footnote{The Wow!\ signal is especially difficult to follow up with high-gain antennas because of the poor source localization.} To observe a signal once or twice requires approximately as much dwell time on the target as the reciprocal of the repetition rate. \citet{howard04} offers a treatment of upper limits along this axis in the context of a pulsed optical SETI program.

Today, most radio telescopes are sensitive to both orthogonal polarizations, and so would detect signals of any polarization, but historically some past searches have only been sensitive to a single circular polarization, and so the sensitivity function for these searches needs polarization as a parameter. \citet{Tarter10} used two polarization dimensions, presumably to capture two of the degrees of freedom described by the Stokes parameters for a purely polarized signal of given intensity $I$.

The final dimension, modulation, is not really a haystack dimension but a catch-all for the nature of the signal being sought. The most obvious forms of modulation are something like AM or FM encodings of information, but this dimension would also include, for instance, a drifting frequency due to the mutual acceleration of the transmitter or receiver or some more complicated mode of modulation. Indeed, a given telescope, instrumentation, and search algorithm might record the appropriate photons and yet still miss a signal if, for instance, the signal were buried in an unmeasured form \citep[for instance, as photon orbital angular momentum;][]{POAM} or had unexpected spectral signature (for instance, as a narrowband signal of rapidly and/or irregularly varying frequency beyond the limits of the search algorithm). This dimension might best be thought of as a line along which many discrete eight-dimensional haystacks sit.

For an instrument and algorithm capable of identifying a given set of signal modulations, then, we can define our haystack of the other eight dimensions, and accept that some forms of signals (including some extreme or unanticipated modulations) will be missed, and so sit outside the haystack.  

The volume of the haystack is then an definite volume integral in this 8D space, and the fraction searched can be calculated given the sensitivity function for a given survey.

\subsection{Haystack Boundaries}

Calculating search completeness requires defining the boundaries of the search; in many cases, one can always look harder for ever rarer or more subtle or more extreme events. The huge range of potential technosignatures also means that one survey cannot hope to be sensitive to all of them, requiring searchers to hypothesize a particular form of signature that they can put interesting limits on with their survey.

For instance, the Drake figure of merit in Section~\ref{SSFM} implicitly defined a haystack with a boundary at a certain distance; while searches would be sensitive to signals from beyond that distance (indeed, out to almost {\it any} distance if they were strong enough) survey speed (or, in our case, completeness) would only be calculated with respect to signals up to that threshold distance. Similarly, with respect to sensitivity: while searches would be extremely sensitive to nearby sources, the figure of merit only considered sensitivity to signals stronger than some threshold.

Examples of haystack boundaries that might define a ``complete'' survey would be:
\begin{enumerate}
\item All transmissions above a threshold power, or all signals above a certain received flux.
\item All transmissions in a range of frequencies or (more practically) all frequencies to which a particular receiver or instrument is sensitive.  In the radio, anthropogenic interference may also pose a practical limit here, meaning one must be content to search a complex set of ranges of frequencies.
\itemrange{2} All transmissions that originate within a given distance, or in a given fraction of the sky, or from a set of sources (e.g.~all stars within 10 pc, all Sun-like stars in the Northern Hemisphere brighter than V = 10, etc.)
\item Transmissions narrower than some bandwidth.  For narrowband searches, this upper boundary might be set by the signal detection algorithm (which might only search for narrow-band signals) and the nature of the data reduction (which might produce power spectra with the wrong resolution to identify the signal).
\item Transmissions with characteristic repetition rates below some threshold.  For instance, a single visit of five minutes' duration would be sensitive to typical repetition periods five minutes or less (including continuous transmissions).  Making 288 such visits would rule out transmissions with typical repetition periods of a day or less, etc.
\item Signals of any polarization strength or form (including unpolarized signals).
\item All signals of a given modulation.

\end{enumerate}

\section{An Example Haystack Calculation}
\subsection{Defining the axes and boundaries}
To illustrate a haystack calculation, we have chosen a particular set of haystack dimensions that might exemplify any particular narrowband radio search with an eye toward determining whether we can yet rule out large numbers of narrowband transmitters in the Solar neighborhood (that is, determining a rigorous---if weak---upper limit for SETI).

Where possible, we choose our dimensions in terms of the properties of the transmitter. Following Drake, we define a complete search as one that is sensitive to all transmitters more powerful than some threshold within a given volume at a given time. This means that, when calculating a search completeness, we will have to exclude the search space volume of a given observation if it lies beyond our haystack boundaries; for instance, of nearby transmitters with powers beyond our chosen threshold.  

\subsection{The Nine Dimensions}
\subsubsection{Sensitivity}

\label{sensitivity}
We define our sensitivity dimension $S$ as 
\begin{equation}\label{eq:sensitivity}
S = \frac{4\pi d^2}{\rm EIRP} = \phi_{\rm min}^{-1}
\end{equation}
\noindent i.e.~having units of inverse flux (m$^2$ W$^{-1}$). We choose this because it has a nice property: for observations made at maximum spectral resolution, it improves linearly with time for unresolved signals.\footnote{This is because increasing integration time allows for finer channelization of data, and given that unresolved signals will always occupy one channel, the noise will stay constant with time at the system equivalent flux density \citep[see Eq.~2 of][]{Enriquez17}.  \citet{Enriquez17} formulated their survey speed under the assumption of {\it fixed} channelization with respect to integration time, which is how most radio surveys operate in practice; one can show that our expression is linear in time from their Eq.~4 (as corrected in \href{https://arxiv.org/abs/1709.03491v2}{arXiv:1709.03491v2}) by setting $\delta \nu = 1/\tau_{\rm obs}$ and $\phi = S_{\rm min, narrow} \delta \nu_{\rm t}$.  Our choice is thus not linear in time for Breakthrough Listen data products.}

The sensitivity of an instrument is a complex function of the other haystack dimensions. When rigorously computing haystack volumes, it would be useful to compute this function explicitly for each observation, ideally using signal injection and recovery tests that span the haystack of interest. At the very least, surveys should report a rough idea of how their sensitivity changes across them.


Our boundaries in this dimension are zero at the bottom (infinitely strong signals) and some EIRP at the top, making the boundary a function of distance. Following \citet{Enriquez17}, we choose a value equivalent to the Arecibo Planetary Radar EIRP of $\sim 10^{13}$ W (Arecibo has an L-band gain of $\approx 2\times 10^{7}$ and a transmission power of $\approx$ 1 MW). 

\subsubsection{Central frequency}

\label{frequency}
The central transmission frequency is the classic haystack dimension, with instrument bandwidth (${\rm BW}_i$) appearing in every radio communication SETI search-speed and haystack definition we are aware of (even sensitivity does not appear in all of them; see, for instance, \citep{Seeger85,Harp16}) The large width of the terrestrial microwave window compared to typical instrument bandwidths in the 1980s made this one of the most daunting dimensions to thoroughly search; the large bandwidth of modern radio instrumentation (such as the 6 GHz Breakthrough Listen backend \citep{MacMahon18}) means that modern instruments can search the radio SETI haystack orders of magnitude more quickly.

Sensitivity is a complex function of central frequency, incorporating the backend and receiver response, atmospheric background emission, astrophysical sources, and the confounding effects of anthropogenic radio frequency interference. 

For our haystack, we somewhat arbitrarily choose a range of frequencies from near the ionospheric cutoff to the upper frequency range of the current instrumentation at the Green Bank Telescope, from 10 MHz up to 115 GHz.  

\subsubsection{Space (distance and direction)}

In the spirit of the NASA Sky Survey, our haystack is agnostic to where a transmitter would be---we do not favor particular stars---and so we parameterize our spatial dimensions as a literal volume centered on the solar system.

For a lower limit, we choose zero distance (we could also choose the edge of the near field of a fiducial telescope---our exact choice on the bottom makes virtually no difference to the haystack volume.)  Following Drake, we place an upper limit at some distance out to which we wish to be complete, $d_{\rm max}$.  We choose 10 kpc for this value in our example haystack, which is roughly the distance to the center of the galaxy.

\begin{figure}[!htbp]
\centering
\includegraphics[width=0.6\textwidth]{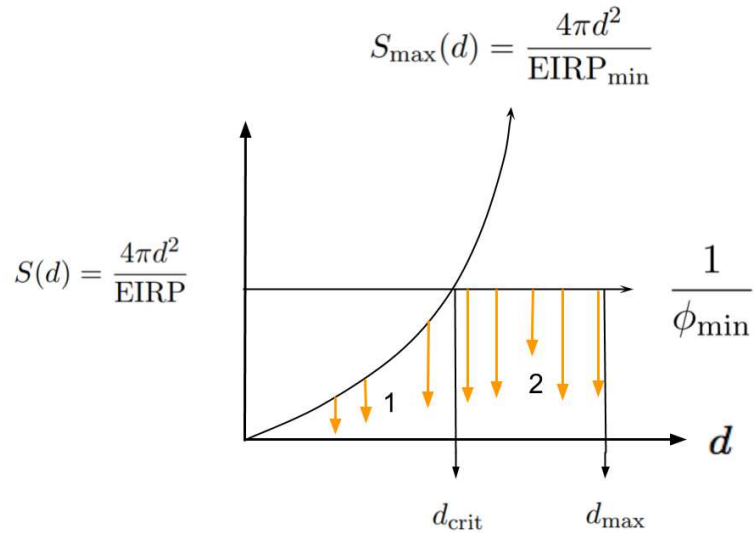}
\caption{\label{fig:distance} Schematic representation of our haystack in $d-S$ (distance sensitivity) space.  At a given sensitivity, we are searching all of the haystack within some distance $d_{\rm crit}$, shown as region 1 (we get no ``credit'' for ruling out signals stronger than our haystack upper limit), and from $d_{\rm crit}$ to $d_{\rm max}$ we are incomplete (region 2).}
\end{figure}

\autoref{eq:sensitivity} gives our sensitivity $S$ an explicit dependence on distance. \autoref{fig:distance} shows how this effects a completeness calculation. The haystack has a parabolic upper surface in the $d-S$ plane, but a given observation is only sensitive to some minimum flux $\phi_{\rm min}$.  Within some critical distance $d_{\rm crit}$ given by $S(d_{\rm crit}) = \phi_{\rm min}^{-1}$ we are more sensitive than necessary to search the haystack---we are still searching straw, but it does not ``count'' toward our completeness because it is part of a larger haystack. Region 1 in the figure shows the volume that does ``count,'' under the parabola.

Beyond $d_{\rm crit}$ we are incomplete, and the volume searched is simply $(d_{\rm max}-d_{\rm crit}) \phi_{\rm min}^{-1}$.

For our other two dimensions, we choose to be complete when we have searched the entire sky, so our boundaries are at zero and $4\pi$ sr.  Our sensitivity in this angular dimension is given by the beam profile of the telescope, and so is properly a function of wavelength and the astrophysical background in each direction; for our calculation we simply approximate this beam as a uniform disk with diameter given by the diffraction limit of the telescope at the center of the band and no astrophysical background.

\subsubsection{Transmission Bandwidth}

The amount of instrument noise competing with a signal of a given strength depends on the bandwidth of the transmission.  Wide transmissions occupy more simultaneous channels, and because each channel contributes noise, we expect that the narrowband signal will be easier to detect than the broadband, for two signals of equivalent flux.  

Extremely broadband signals, such as those produced by very short pulses, may require searches in the time domain (i.e.\ the coherently measured voltages at the telescope) but should still be detectable in the spectral domain as extremely wide increases in power in a given direction.  In this case, the signal may be spread out beyond the limits of the receiver, further reducing a telescope's sensitivity to a signal of a given luminosity.

\begin{figure}[hp]
\centering
\includegraphics[width=0.8\textwidth]{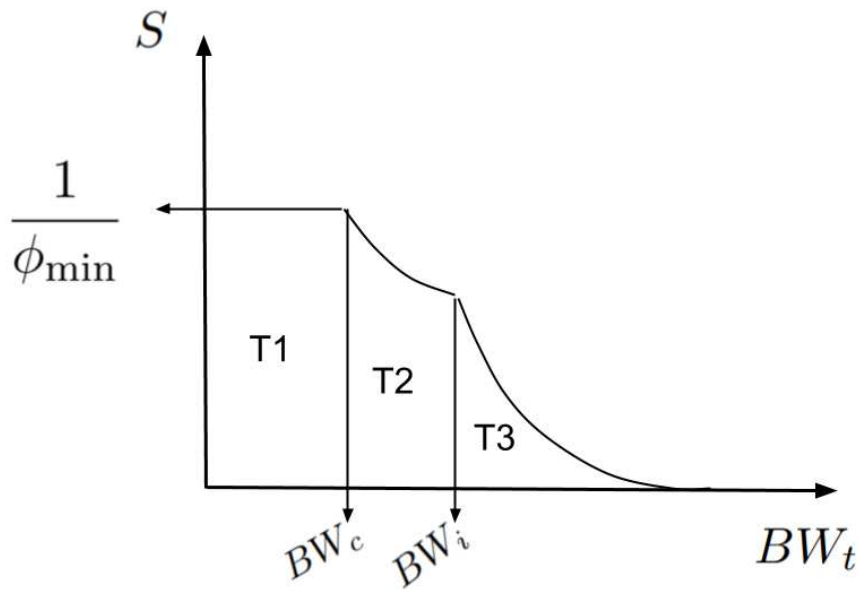}
\caption{\label{fig:tbw} Schematic showing the sensitivity of an instrument to signals of a given flux as a function of the transmission bandwidth. Below the width of a single channel, it is flat because all of the signal falls into a single channel (region T1); beyond this, it drops first because the signal is spread out among more channels, each contributing noise (region T2); and beyond that, it drops because much of the signal lands outside the bandwidth of the instrument and so goes undetected. \label{bandwidth}} 
\end{figure}

In general, then, sensitivity is a decreasing function of transmission bandwidth BW$_t$ as illustrated in Figure~\ref{bandwidth}: it is flat for unresolved signals (those with BW$_t < $BW$_c$), and beyond this it drops as the square root of the number of channels it occupies (each contributing some amount of noise).  Once BW$_t$ exceeds that of the {\it entire instrument} (BW$_i$) it falls inversely with bandwidth. 

For our example haystack, we choose bandwidth boundaries of zero bandwidth at the low end (i.e.~arbitrarily narrow transmissions, although in practice there will be some astrophysical limit due to the solar wind, if nothing else) and a high end of 20 MHz, somewhat arbitrarily chosen to match our the lower bound of the central frequencies in our haystack.

To calculate a haystack volume, we will need to compute a volume integral that includes the term
\begin{equation}
\label{FourVolume}
\iiint S(\nu, d, {\rm BW}_t) \dd{\nu} \dd{d} \dd{{\rm BW}_t}
\end{equation}
\noindent with the upper and lower limits being set by the either the boundary of the haystack or the sensitivity of a particular observation, depending on which region of $(\nu, d, {\rm BW}_t)$ space we are in and how the haystack boundaries have been chosen. The calculation of the resulting 4-volume is complex, but as it turns out, analytic for observations of uniform sensitivity across an instrumental bandpass and the idealized instrumental response to transmission bandwidth shown in Figure~\ref{bandwidth}. While this is an idealized case (see Section~\ref{frequency}) it is still a sufficiently useful approximation that we can at least generate an order-of-magnitude estimate of the searching that has been done, given basic instrumental and algorithmic parameters for a given search.
It must also be noted that this sensitivity estimation assumes that such a top hat signal (in frequency space) would be detected by the algorithms and flagged as being artificial, even if it is broadband. This is because perfect top hat signals do not occur naturally. In reality we do not know the algorithmic responses to various kinds of signals, including broadband signals near our upper limit, for the searches discussed in \autoref{haystacks}, so our estimate is likely quite generous.

We illustrate the analytic calculation for a particular case of Equation~\ref{FourVolume} in the appendices, and include with this paper a Mathematica notebook and Python script that calculate a more general case\footnote{The Mathematica and Python scripts, along with their input files, can be found at \url{https://doi.org/10.5281/zenodo.1409818}.}.

\subsubsection{Repetition Rate}

For the repetition-rate axis, we follow \citet{Gray02} and simply use the total integration time on a target across all observations; our sensitivity is thus a step function at its full value for repetition times shorter than this number, and zero beyond it.  In principle, one could hypothesize a range of transmission repetition schemes and more rigorously calculate the fraction ruled out by the observations to date, but our approach should suffice for the general case of unknown repetition schemes.

In principle, additional observations could be stacked with previous observations to also increase sensitivity to fast repetition rates, but this is really only practical if the observations being stacked are less than a few hours apart, because the accelerations of the telescope and source will eventually become difficult to model (i.e.~they will not be a simple linear Doppler drift). However if one hypothesizes that the transmitter frequency is modulated to be constant in the barycentric frame of the Solar System, one could remove the receiving telescope's acceleration and perform such stacking.  Here, we will assume repeat visits to a source do not improve sensitivity.

For our haystack, we choose an upper boundary equal to a repetition rate of once per year.

\subsubsection{Transmission polarization}\label{sec:polarization}

Following the \citet{Dreher97} figure of merit as adopted by \citet{Tarter01}, we adopt a single parameter $\eta_{\rm pol}$ that collapses the polarization dimensions to a scalar from 0 to 1, representing the fraction of all polarizations to which a given survey is sensitive. We presume modern surveys have $\eta_{\rm pol} = 1$, but historical searches sensitive to only a single polarization (linear or circular) would be assigned 0.5, representing their decreased sensitivity along this axis.

\subsubsection{Modulation}

This catch-all dimension is the most difficult to quantify. In terms of the amplitude and frequency modulation of a radio signal, the ideal way to incorporate it is to evaluate one's signal detection algorithms for injected signals of a wide variety of modulations.  One could then define the haystack one is searching in terms of those modulations to which one is essentially completely sensitive above some EIRP, or include additional modulation dimensions along which one's sensitivity is variable.  Here, we define our haystack's modulation to be signals of constant strength across some constant transmission bandwidth and with central frequency unmodulated except for some modest Doppler drift (on the order of the Earth's barycentric acceleration).  Such a radio survey would also be sensitive to more complex modulations, naturally, but those would be outside of our haystack.

\subsection{Some Example Search Volumes}
Our entire haystack volume can be expressed as (see Equation~\ref{totalHaystack}):
\begin{equation}
V_{\rm haystack} = \frac{4\pi\Omega_{\rm max}d_{\rm max}^5}{5\ {\rm EIRP_{min}}}{\rm BW}_{t,{\rm max}} (\nu_{\rm max} - \nu_{\rm min}) T_{\rm max}
\end{equation}
\noindent where $\nu$ refers to the central frequency of the transmission and $T$ is the maximum repetition period. Recapping the above, our example haystack boundaries are:

\begin{tabular}{|l|lr|}
\hline
Dimension & Lower Bound & Upper Bound \\
\hline
Sensitivity (expressed here as EIRP) & $10^{13}$ W & $\infty$ \\
Distance ($d$)& 0 & 10 kpc \\
Solid angle ($\Omega$) & 0 & 4$\pi$ sr\\
Transmission central frequency ($\nu$) & 10 MHz & 115 GHz \\
Transmission bandwidth (BW$_t$) & 0 & 20 MHz \\ 
Repetition Period ($T$) & Continuous & 1 year\\
Polarization fraction & 0 (Unpolarized) & 1 (Completely polarized) \\
\hline
\end{tabular}

This leads to a total 8D haystack volume of $6.4 \times 10^{116}$ m$^5$Hz$^2$sW$^{-1}$ (recall that we express the sensitivity axis in units of m$^2$W$^{-1}$).  

From the Breakthrough Listen L-band campaign described by \citet{Enriquez17}, we approximate the following search parameters in these dimensions:

\begin{tabular}{|l|l|}
\hline
Parameter & Value \\ 
\hline
Single-channel Sensitivity ($S/{\rm BW}_c$) & 17 Jy \\
Channel bandwidth (BW$_c$) & 2.7 Hz \\
Instrument bandwidth (BW$_i$) & 800 MHz (1.1--1.9 GHz)\\
Solid angle surveyed ($\Omega$) & 10.6 sq.\ deg.\ ($3.2\times10^{-3}$ sr) \\
Transmission bandwidth sensitivity ($S({\rm BW}_t))$ & Uniform up to at least 20 MHz \\ 
Total time on targets ($T$) & 300s \\
Polarization sensitivity & 1 (both senses) \\
\hline
\end{tabular}

We use 17 Jy as the single channel sensitivity because \citet{Enriquez17} used a 25$\sigma$ threshold for detection and 300s for the total time on target, even though most targets were visited three times for this duration because they required the signal to be present during all three ``on'' visits to be a detection.  We assume for this model that even transmissions with bandwidths as wide as 20 MHz would have been detected, even though it is not clear whether this is true for the actual algorithm used for this survey. The total area surveyed is approximated simply as the beam size at 1.5 GHz times the number of targets. Finally, much of the L-band region is contaminated with radio frequency interference, so some small but non-negligible portion of this survey has much worse sensitivity than we have assumed. Certainly, this haystack volume can be calculated more precisely for this survey, but this approximation should be sufficient for our purposes here.

This amounts to a total searched space for the 692 stars of $2.4\times10^{98}$ m$^5$Hz$^2$sW$^{-1}$, and the haystack fraction up to $3.8\times 10^{-19}$.  

We similarly compute the search fraction of all of the searches reported in Tables 3 of \cite{Enriquez17} in Table~\ref{haystacks}, plus a few others. The largest surveys by our haystack metric are the MWA surveys of \citet{tingay2018} and \citet{tingay2016},\footnote{Although the methodology sections of these papers highlights only a targeted search towards a few exoplanet host stars, these papers' upper limits include the entire primary beam of the telescope because the entire data cube was searched for narrow band emission (S.~Tingay, private communication).} although the Breakthrough Listen survey is catching up. The sum of all of these searches comes out to $6.0 \times 10^{-18}$.

\begin{table}[]
\centering
\caption{8D Haystack search fractions for some radio SETI searches \label{haystacks}}
\resizebox{\columnwidth}{!}{
\begin{tabular}{lllccccccccc}
\hline \hline
 & Instrument              & Survey                                     & T\footnote{Time on each target}     & BW$_c$\footnote{Single channel bandwidth } & F$_{\nu , \rm{min}}$\footnote{Minimum survey sensitivity (specific flux): Threshold $\times$ 1-$\sigma$ Sensitivity} & $\phi_{\rm{min}}$\footnote{Specific Flux $\times$ Channel BW} & Beamwidth & $\nu_{\rm max}$ & $\nu_{\rm min}$ & Sky Coverage & Haystack fraction   \\ 
 &                         &                                            & (s) & (Hz) & (Jy) & (10$^{-26}$ W/m$^2$)               & (arcmin)  & (GHz)           & (GHz)           & (deg$^2$)              &             \\ \hline
 & GBT 100 m               & Breakthrough Listen\footnote{\cite{Enriquez17}}         & 300   & 2.7    & 17 & 46             & 8         & 1.9             & 1.1             & 10.6            & 3.8 $\times 10^{-19}$ \\
 & VLA                    & \cite{Gray2017}           & 1200  & 122   & 0.24 & 29      & 36        & 1.401            & 1.399           & 2.25\footnote{The entire area searched was 1 deg$^2$, but sensitivity near the field edge was low. For consistency with the other surveys, we approximate the sky coverage as being constant across a disk with diameter equal to 1.22 $\lambda$/D}             & 1.5 $\times 10^{-20}$ \\
 & ATA (HZ Exoplanet)     & \cite{Harp16_exoplanets} & 93    & 0.7    & 378   & 265          & 4.24      & 9.00            & 1.00            & 0.26             & 2.5 $\times 10^{-21}$ \\
 & ATA (Exoplanet, not HZ)& \cite{Harp16_exoplanets} & 93    & 0.7    & 378   & 265          & 4.24      & 6.000           & 4.00            & 7.7           & 1.9 $\times 10^{-20}$ \\
 & ATA (HabCat)           & \cite{Harp16_exoplanets} & 93    & 0.7    & 378   & 265          & 4.24      & 5.168           & 4.832           & 11           & 4.6 $\times 10^{-21}$ \\
 & ATA (Tycho)            & \cite{Harp16_exoplanets} & 93    & 0.7    & 378    & 265         & 4.24      & 5.134           & 4.866           & 29           & 9.6 $\times 10^{-21}$ \\
 & GBT (100m)              & \cite{Siemion2013}        & 300   & 1      & 10 & 10              & 8         & 1.9             & 1.1             & 1.3             & 1.3 $\times 10^{-19}$ \\
 & Arecibo                 & Project Phoenix\footnote{\label{phoenixAO}\cite{Backus2002}}                            & 276   & 1      & 16 & 16             & 3         & 1.75            & 1.2             & 0.58            & 2.3 $\times 10^{-20}$ \\
 & Arecibo                 & Project Phoenix$^{\rm\ref{phoenixAO}}$                            & 195   & 1      & 16   & 16         & 2         & 3               & 1.75            & 0.33            & 2.1 $\times 10^{-20}$ \\
 & Parkes                  & Project Phoenix\footnote{\label{Parkescitation}\cite{Backus1995}; \cite{Tarter1996}; \cite{Backus1997}; \cite{Backus1998}}                            & 276   & 1      & 100   &100         & 13        & 1.75            & 1.2             & 7.85            & 5.0 $\times 10^{-20}$ \\
 & Parkes                  & Project Phoenix$^{\rm\ref{Parkescitation}}$                            & 138   & 1      & 100   & 100         & 8         & 3               & 1.75            & 1.46            & 1.0 $\times 10^{-20}$ \\
 & NRAO 140 ft             & Project Phoenix\footnote{\cite{Cullers2000}}                            & 552   & 1      & 100      & 100         & 14        & 3               & 1.2             & 8.3            & 3.4 $\times 10^{-19}$ \\ 
  & MWA             & \cite{tingay2016}                            & 6900 & $10^4$      & 4.9\footnote{For rough consistency with the other surveys we assign the sensitivity here to be 7 $\times$ 1-$\sigma$ upper limits given in the paper. \label{tingay}}  & 49000             & 1800 & 0.135               & 0.104             & 625          & 0.72 $\times 10^{-18}$ \\
    & MWA             & \cite{tingay2018}                           & 10800   & $10^4$      & 2$^{\rm \ref{tingay}}$  & 20000             & 1800        & 0.130              & 0.0992          & 400          & 4.28 $\times 10^{-18}$ \\
  \hline
\end{tabular}
}
\end{table}

The total volume of Earth's oceans is approximately $1.335\times10^{21}$ liters \citep{OceanVolume}, meaning that the total searching done to date is equivalent of $\sim$8000 l of seawater, which is somewhere between the volumes of a large hot tub and a small swimming pool. This is significantly larger than the \citet{Tarter10} estimate of one drinking glass, but \citeauthor{Tarter10} used a different haystack definition and many of the surveys in our calculation were performed after the \citeauthor{Tarter10} estimate.  Even our larger estimate underscores how little searching has actually occurred. One hopes that the Cosmic Haystack is rich with needles.  

We reemphasize, however, that our numbers are very sensitive to our haystack definition and boundaries (recall that our haystack volume goes as $d_{\rm max}^5$), and that searches such as \citet{Enriquez17} and \citet{Harp16_exoplanets} are {\it much} more complete with respect to the number of nearby stars searched for continuous, narrowband transmitters.

\section{Conclusions}

We have demonstrated one way to rigorously define both an $n$-dimensional haystack and define the fraction of it searched for alien ``needles'' by a given SETI survey. For our example haystack of radio transmissions, we have made a rough calculation of the total fraction of it searched to date under the (incorrect but sufficient for our purposes) assumptions that published surveys have comparable search strategies and uniform sensitivities across their instrumental bandpasses and are uniformly sensitive to narrow and broadband transmissions. We show that these assumptions allow us to do our calculation analytically, and we get a result very similar to that of another haystack, as calculated by \citet{Tarter10}: our current search completeness is extremely low, akin to having searched something like a large hot tub or small swimming pool's worth of water out of all of Earth's oceans.  

Such calculations should become routine for those who wish to calculate the completeness of a SETI program, or who would like to put quantitative upper limits on the existence of technosignatures.  A more rigorous haystack search fraction (or parameter space upper limit) for a given search for radio transmissions might be computed by determining the sensitivity of a survey $S$ to the other haystack parameters via signal injection and recovery methods, and performing a numerical volume integral, rather than the approximate, analytic formula here.

However, perhaps more importantly for now, even order-of-magnitude estimates of the SETI space searched to date are valuable because they rebut the pervasive misconception that SETI work to date has significantly ``sharpened'' the Fermi Paradox or proven so dispositive that SETI can be said to have ``failed'' to find what it seeks.  

We should be careful, however, not to let this result swing the pendulum of public perceptions of SETI too far the other way by suggesting that the SETI haystack is so large that we can never hope to find a needle. The whole haystack need only be searched if one needs to prove that there are zero needles---because technological life might spread through the Galaxy, and/or technological species might arise independently in many places, we might expect there to be a great number of needles to be found. Also, our haystack definition included vast swaths of interstellar space where we have no particular reason to expect to find transmitters; humanity's completeness to subsets of this haystack---for instance, for continuous, permanent transmissions from nearby stars---is many orders of magnitude higher.

Both completeness and upper limit calculations are close conceptual cousins of survey figures of merit, which typically include bandwidth, sky coverage, and sensitivity among their terms.  In most cases, it is clear that surveys with large bandwidth, wide fields of view, long exposures, repeat visits, and good sensitivity can allow for orders-of-magnitude faster searches of haystacks than surveys without these qualities. This is illustrated nicely by the \citeauthor{tingay2018} surveys, which dominate our haystack search volume with only a few hours of integration time because of their extremely large sky coverage. The dream of ``all-sky, all the time'' high-bandwidth surveys remains worth pursuing.\footnote{Along these lines, we note the PANO-SETI program of Shelley Wright and Paul Horowitz, Laser SETI by Eliot Gillum, and efforts at the SKA and other arrays (\url{http://www.planetary.org/blogs/jason-davis/2017/20171025-seti-anybody-out-there.html}) to be discussed at the upcoming meeting at Jodrell Bank (\url{http://www.jodrellbank.manchester.ac.uk/news-and-events/wide-field-seti-workshop/}).}

\appendix

For brevity, we use the following notation in this appendix:
\begin{enumerate}
\item $P_0 \equiv \frac{\rm EIRP_{\rm min}}{4\pi} $.
\item $S_{\rm max} \equiv \frac{d^2}{P_0} $ as shown in Figure \ref{fig:distance}.
\item $ c \equiv BW_{c}$.
\item $ i \equiv BW_{i}$.
\item $ t \equiv BW_{t}$.
\end{enumerate}

In this appendix, we present the derivation for the haystack volume searched for the special case of a search with the properties:
\begin{enumerate}
\item $t_{\rm max} > i > c > t_{\rm min} > 0$
\item $d_{\rm max} > d_{\rm crit} > 0$
\item $\nu_{\rm min,hackstack} + {\rm BW}_{t,{\rm max}} < \nu_{\rm low}$
\end{enumerate}
\noindent where $\nu_{\rm low}$ represents the lowest frequency measured by a given survey and $\nu_{\rm min,haystack}$ refers to the haystack lower boundary.

\section{Integration over central frequency}

\begin{center}
 General definition for $d > d_{\rm crit}$
\end{center}

\begin{figure}[h]
\centering
\includegraphics[width=0.5\textwidth]{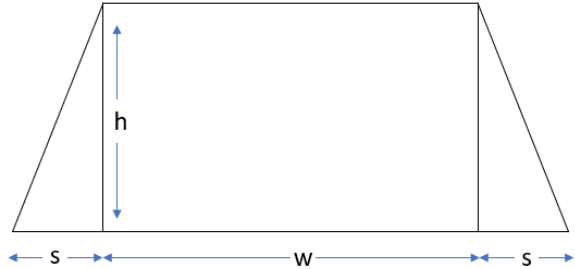}
\caption{\label{fig:trapezium} Showing the general framework we have used to help integrate over the central frequency for the faraway case. The entire trapezoid shown is included in the haystack. The x-axis represents central transmission frequency and the y-axis is the sensitivity. Here, $h$, $s$, and $w$ are dummy variables.}
\end{figure}

We calculate the haystack volume as an $n$-dimensional integral under the sensitivity $S$ as a function of the other dimensions. The first dimension we choose to integrate along is the central frequency of transmission. In Figure \ref{fig:trapezium} we schematically illustrate an idealized sensitivity function that is uniform at central frequencies within the instrumental bandpass, and then worsens linearly with frequency for central frequencies and transmission bandwidths that have some of the signal falling outside the instrumental bandpass. The area of this shape is simply

\begin{subequations} \label{eq:Ai1}
\begin{align}
A_{i1} &= w  h + 2 \times \left(\frac{1}{2} s  h  \right) \\
&= h(w+s)
\end{align}
\end{subequations}

\begin{center}
 General definition for d $\leq$ dcrit
\end{center}

\begin{figure}[htbp]
\centering
\includegraphics[width=0.5\textwidth]{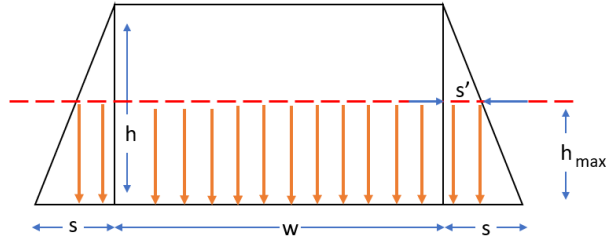}
\caption{\label{fig:trapezium_ceilingdrop} Showing the general framework we have used to help integrate over the central frequency for the case of transmitters closer than the critical distance $d_{\rm crit}$.  In this case, the haystack upper boundary at $h_{\rm max}$ cuts off our volume---the search is sensitive to weaker transmitters, but these lie outside our haystack so we do not count them.}
\end{figure}

In the case for objects closer than $d_{\rm crit}$ the haystack upper boundary S$_{\rm max}$ is lower than our survey sensitivity $S$, and so reduces the area under the curve as shown in Figure \ref{fig:trapezium_ceilingdrop}.

Here, 
\begin{equation}
s' = s \frac{h_i - h_{\rm max}}{h_i}
\end{equation}

\begin{subequations}\label{eq:Ai2}
\begin{align}
A_{i2} &= A_{1i} - (h_i - h_{\rm max})\left( w + s' \right) \\
&= A_{i1} - \left(1 - \frac{h_{\rm max}}{h_i} \right) \left\lbrace h_i (w+s) - h_{\rm max}s \right\rbrace
\end{align}
\end{subequations}

\subsection{Spectrally unresolved signals: t $<$ c}
In this case, the out-of-band portion of the sensitivity is given by signals with central frequencies closer than $t/2$ to the instrumental bandpass boundary.  So we have
\begin{equation}
w = i-t
\end{equation}
\begin{equation}
s = t
\end{equation}
and the sensitivity, in this case $S_1$ is simply the survey sensitivity, so we have:
\begin{equation}
h = S_1 = \frac{1}{\phi_{\rm min}} 
\end{equation}

The area under the curve for distant sources with $d > d_{\rm crit}$ is 
\begin{subequations} \label{eq:A11}
\begin{align}
A_{11} &= h (w+s) \\
&= \frac{i}{\phi_{\rm min}}
\end{align}
\end{subequations}

For the case of nearby sources with $d \leq d_{\rm crit}$ we additionally assign $h_{\rm max} = S_{\rm max}$ and so we have
\begin{subequations}\label{eq:A12}
\begin{align}
A_{12} &= A_{11} - \left(1 - \frac{S_{\rm max}}{S_1} \right) \left\lbrace S_1i - S_{\rm max}t \right\rbrace \\
&= S_{\rm max} \left\lbrace i + t(1-\phi_{\rm min} S_{\rm max}) \right \rbrace
\end{align}
\end{subequations}

\subsection{Spectrally resolved signals: $i > t \geq c$}

As in the previous case, we have

\begin{equation}
w = i-t
\end{equation}

\begin{equation}
s = t
\end{equation}

\noindent but now our sensitivity is given by

\begin{equation}
h = S_2 = \frac{1}{\phi_{\rm min}} \sqrt[]{\frac{c}{t}} 
\end{equation}

For distant sources with $d > d_{\rm crit}$ area is
\begin{subequations}\label{eq:A21}
\begin{align}
A_{21} &= h (w+s) \\
&= \frac{1}{\phi_{\rm min}} \sqrt[]{\frac{c}{t}} i
\end{align}
\end{subequations}
\noindent and for nearby sources with $d \leq d_{\rm crit}$ we have
\begin{subequations}\label{eq:A22}
\begin{align}
A_{22} &= A_{21} - \left(1 - \frac{S_{\rm max}}{S_2} \right) \left\lbrace S_2i - S_{\rm max}t \right\rbrace \\
&= S_{\rm max} \left( i + t - S_{\rm max}\phi_{\rm min} \sqrt[]{\frac{t^3}{c}} \right)
\end{align}
\end{subequations}

\subsection{Very broadband signals: $t \geq i > c$}

In this case, the sensitivity drops linearly with transmission bandwidth
\begin{equation}
h = S_3 = \frac{1}{\phi_{\rm min}}\frac{\sqrt[]{ci}}{t}
\end{equation}
\noindent and now the base of our trapezoid has a width equal to the transmission bandwidth, not the instrument, so we have
\begin{equation}
w = t - i 
\end{equation}
\begin{equation}
s = i
\end{equation}

Therefore, area under the curve for distant sources with $d > d_{\rm crit}$ is
\begin{subequations}\label{eq:A31}
\begin{align}
A_{31} &= h (w+s) \\
&= \frac{\sqrt[]{ci}}{\phi_{\rm min}}
\end{align}
\end{subequations}
\noindent and for nearby sources with $d \leq d_{\rm crit}$ we have
\begin{subequations}\label{eq:A32}
\begin{align}
A_{32} &= A_{31} - \left(1 - \frac{S_{\rm max}}{S_3} \right) ( S_3t - S_{\rm max}i) \\
&= S_{\rm max} \left \lbrace i + t\left(1  -  S_{\rm max}\phi_{\rm min} \sqrt[]{\frac{i}{c}} \right) \right \rbrace 
\end{align}
\end{subequations}

\section{Integrating over transmitter bandwidth and distance}

We now have defined six regions, $A_{ij}$ which have units of sensitivity times (central) frequency and which are functions of transmitter bandwidth $t$. We now integrate the area under these functions as a function of $t$, weighted by the volume of space they explore, (i.e.~weighting them by $d^2$). 

\subsection{Spectrally unresolved signals: $t < c$}

Our areas are taken from Equations~\ref{eq:A11} \& \ref{eq:A12}. In this case, $d_{\rm crit} = \sqrt{\frac{P_o}{\phi_{\rm min}}}$.

For $d>d_{\rm crit}$
\begin{subequations}\label{eq:V11}
\begin{align}
V_{11} &= \int_{0}^{c}\dd{t} \int_{d_{\rm crit}}^{d_{\rm max}}\dd{d}A_{11} d^2 \\
&= \int_{0}^{c}\dd{t} \int_{d_{\rm crit}}^{d_{\rm max}}\dd{d} \frac{i d^2}{\phi_{\rm min}}  \\
&= \frac{ci}{3\phi_{\rm min}} \left(d_{\rm max}^3 - \left(\frac{P_o}{\phi_{\rm min}}\right)^{3/2}\right)
\end{align}
\end{subequations}

For $d<d_{\rm crit}$

\begin{subequations}\label{eq:V12}
\begin{align}
V_{12} &= \int_{0}^{c}\dd{t} \int_{d_{\rm min}}^{d_{\rm crit}}\dd{d}A_{12} d^2 \\
&=\int_{0}^{c}\dd{t} \int_0^{d_{\rm crit}}\dd{d} S_{\rm max}\left(i+ t(1-\phi_{\rm min} S_{\rm max})\right) d^2 \\
&= \frac{1}{5P_0}\left(\frac{P_0}{\phi_{\rm min}}\right)^{\frac{5}{2}} \left(ic+\frac{c^2}{2}\right) - \frac{1}{7 P_{0}^2}\left(\frac{P_0}{\phi_{\rm min}}\right)^{\frac{7}{2}}\frac{c^2\phi_{\rm min}}{2}
\end{align}
\end{subequations}

\subsection{Spectrally resolved signals: $i > t > c$}
Our areas are taken from Equations \ref{eq:A21} and \ref{eq:A22}. In this case, $d_{\rm crit} = \sqrt{\frac{P_o}{\phi_{\rm min}}} \left(\frac{c}{t}\right)^{\frac{1}{4}}$.

For $d>d_{\rm crit}$

\begin{subequations}\label{eq:V21}
\begin{align}
V_{21} &= \int_{c}^{i}\dd{t} \int_{d_{\rm crit}}^{d_{\rm max}}\dd{d} A_{21}  d^2 \\
&= \int_{c}^{i}\dd{t} \int_{d_{\rm crit}}^{d_{\rm max}}\dd{d} \frac{i}{\phi_{\rm min}} \sqrt{\frac{c}{t}} d^2 \\
&= \frac{i}{3 \phi_{\rm min}} \sqrt{c} \int_{c}^{i} \frac{\dd{t}}{\sqrt{t}} (d_{\rm max}^3 - d_{\rm crit}^3) \\
&= \frac{i \sqrt{c}}{3 \phi_{\rm min}} \left( 2 d_{\rm max}^3 (\sqrt{i}-\sqrt{c} + 4\left(\frac{P_0}{\phi_{\rm min}}\right)^{\frac{3}{2}}\left(\frac{c^3}{i}^{\frac{1}{4}} - \sqrt{c}\right)\right)
\end{align}
\end{subequations}

For $d<d_{\rm crit}$

\begin{subequations}\label{eq:V22}
\begin{align}
V_{22} &= \int_{c}^{i}\dd{t} \int_{0}^{d_{\rm crit}}\dd{d}A_{22}  d^2\\
&= \int_{c}^{i}\dd{t} \int_{0}^{d_{\rm crit}}\dd{d} \left[\frac{d^2 i }{P_0} + \frac{d^2 t }{P_0}  - \frac{d^4}{P_0^2}\phi_{\rm min} t \sqrt{\frac{t}{c}}                \right]   d^2 \\
&= \int^{i}_{c} \dd{t} \left( \frac{P_0}{\phi_{\rm min}}\right)^{\frac{3}{2}}\frac{c^{\frac{5}{4}}}{\phi_{\rm min}}\left\lbrace \frac{i t^{-\frac{5}{4}}}{5} + \frac{t^{-\frac{1}{4}}}{5} - \frac{t^{-\frac{1}{4}}}{7}\right\rbrace \\
&= \frac{3}{35} \frac{P_0^{\frac{3}{2}}}{\phi_{\rm min}^{\frac{5}{2}}} \left\lbrace 84ic - 76i^{\frac{3}{4}}c^{\frac{5}{4}} - 8c^2\right\rbrace
\end{align}
\end{subequations}

\subsection{Very broadband signals: $t \geq i > c$}
Our areas are now taken from Equations \ref{eq:A31} and \ref{eq:A32}. In this case, $d_{\rm crit} = \sqrt{\frac{P_o}{\phi_{\rm min}}} \left(\frac{ci}{t^2}\right)^{\frac{1}{4}}$.

For $d>d_{\rm crit}$

\begin{subequations}\label{eq:V31}
\begin{align}
V_{31} & = \int_{i}^{t_{\rm max}}\dd{t} \int_{d_{\rm crit}}^{d_{\rm max}}\dd{d} A_{31}  d^2 \\
&= \int_{c}^{i}\dd{t} \int_{d_{\rm crit}}^{d_{\rm max}}\dd{d} \frac{\sqrt[]{ci}}{\phi_{\rm min}} d^2 \\
&= \frac{\sqrt[]{ci}}{3\phi_{\rm min}} \int_i^{t_{\rm max}} \dd{t} \left[ d_{\rm max}^3 - \left( \frac{P_0}{\phi_{\rm min}}\right)^\frac{3}{2} (ci)^{\frac{3}{4}}t^{-\frac{3}{2}} \right]\\
&= \frac{\sqrt[]{ci}}{3\phi_{\rm min}}d^3_{\rm max} (t_{\rm max} - i) - \frac{P_0^{\frac{3}{2}}}{\phi_{\rm min}^{\frac{5}{2}}} \frac{2(ci)^{\frac{5}{4}}}{3} \left(\frac{1}{\sqrt[]{i}} - \frac{1}{\sqrt[]{t_{\rm t_{\rm max}}}} \right)
\end{align}
\end{subequations}

For $d<d_{\rm crit}$

\begin{subequations}\label{eq:V32}
\begin{align}
V_{32} &= \int_{i}^{t_{\rm max}}\dd{t} \int^{d_{\rm crit}}_0 \dd{d} A_{31}  d^2 \\
&= \int_{i}^{t_{\rm max}}\dd{t} \int^{d_{\rm crit}}_0 \dd{d} \left[ S_{\rm max} \left\lbrace i +t - S_{\rm max}\phi_{\rm min} t  \sqrt{\frac{i}{c}}\right\rbrace \right] d^2 \\
&= \int_{i}^{t_{\rm max}}\dd{t} \left[ \frac{d_{\rm crit}^5}{5P_0}i + \frac{d_{\rm crit}^5}{5P_0}t - \frac{d_{\rm crit}^7}{7P_0^2}\phi_{\rm min} \sqrt[]{\frac{i}{c}}t \right] \\
&= \frac{P_0^{\frac{3}{2}}}{\phi_{\rm min}^{\frac{5}{2}}} 2 (ci)^{\frac{5}{4}} \left\lbrace \left(\sqrt[]{\frac{1}{i}} - \sqrt[]{\frac{1}{t_{\rm max}}} \right) + \frac{1}{3} \left( \sqrt[]{\frac{1}{i^3}} - \sqrt[]{\frac{1}{t_{\rm max}^3}}\right) (i - \sqrt[]{ci})           \right\rbrace
\end{align}
\end{subequations}

\section{The 6D Integral}

The integral over the first six dimensions---space (3), central frequency (1), transmission bandwidth (1), and sensitivity (1) --is the sum of the volumes from Equations \ref{eq:V11}, Equation \ref{eq:V12}, Equation \ref{eq:V21}, Equation \ref{eq:V22}, Equation \ref{eq:V31}, and Equation \ref{eq:V32} times $\Omega$ (the solid angle of the observations).

\begin{equation}
V_{\rm scanned6D} = (V_{11} + V_{12} + V_{21} + V_{22} + V_{31} + V_{32}) \times \Omega
\end{equation}

We can compute the total volume of the 6D haystack as

\begin{subequations}
\begin{align}
V_{\rm total6D} &= \int_0^{4\pi}\int_{\nu_{\rm min}}^{\nu_{\rm max}}\int_0^{BW_{t,{\rm max}}}\int_0^{d_{\rm max}}  d^2 S_{\rm max}(d) \dd{d}  \dd{t} \dd{\nu}\dd{\Omega}\\
&=\frac{d_{\rm max}^5}{5 P_0} \times t_{\rm max} \times (\nu_{\rm max} - \nu_{\rm min}) \times 4\pi
\end{align}
\end{subequations}

\section{8D Integral}
We further include the repetition rate and polarization fraction to obtain the final 8D volume. Here, $\eta_{\rm pol}$ is the polarization fraction as discussed in Section \ref{sec:polarization} and $T_{\rm max}$ is the longest repetition period in the haystack:

\begin{equation}
V_{\rm scanned8D} = V_{\rm scanned6D} \times T_{\rm scan} \times \eta_{\rm pol}
\end{equation}

\begin{equation}\label{totalHaystack}
V_{\rm total8D} = V_{\rm total6D} \times T_{\rm max}
\end{equation}

We provide electronically with this article a Mathematica notebook in which we analytically calculated these integrals. We also provide a Python script that computes numerical values for the total haystack volume and search fraction, which we used to calculate the numbers in Table~\ref{haystacks} and which handles the more general cases where the haystack boundaries have $d_{\rm max} < d_{\rm crit}$ and BW$_{t,{\rm max}} < {\rm BW}_i$ (although we still assume our third condition, that $\nu_{\rm min,hackstack} + {\rm BW}_{t,{\rm max}} < \nu_{\rm low}$, holds.)

\acknowledgements
This paper grew out of a final project in the 2018 graduate course in SETI at Penn State.\footnote{\url{https://sites.psu.edu/seticourse/9-dimensional-haystack/}} The authors are extremely grateful to Dr.~Jill Tarter for suggesting this project and for a careful read of the manuscript, and to the anonymous referee for a prompt, positive, and constructive report that improved the paper.
J.T.W.~thanks Bez Thomas for helping to track down the Spanish equivalent of the the phrase ``needle in haystack'' in {\it Don Quixote}; and Stephen Wilson for help tracking down early and apparently first printed references to the terms ``Cosmic Haystack'' and ``technosignatures.'' S.K. thanks Emilio Enriquez from Breakthrough Listen for the discussions and insight into the sensitivity calculations.


\end{document}